\documentclass[12pt,letterpaper]{article}
\usepackage[latin1]{inputenc}
\usepackage{amsmath}
\usepackage{amsfonts}
\usepackage{amssymb}
\usepackage{epsfig}
\usepackage{a4wide}
\usepackage{authblk}

\begin{document}

\title{Minimal Supersymmetric Hybrid Inflation, Flipped~SU(5) and Proton Decay}

\author{Mansoor Ur Rehman}
\author{Qaisar Shafi}
\author{Joshua R. Wickman\footnote{E-mail address: jwickman@udel.edu}}

\affil{Bartol Research Institute, Department of Physics and Astronomy, 
University of Delaware, Newark, Delaware 19716, USA}

\date{December 2009}


\maketitle

\abstract{
Minimal supersymmetric hybrid inflation utilizes a canonical K\"ahler potential and a
renormalizable superpotential which is uniquely determined by imposing a U(1) $R$-symmetry. In computing the scalar spectral index $n_s$ we take into account
modifications of  the tree level potential caused by radiative and supergravity corrections, as well as contributions
from the soft supersymmetry breaking terms with a negative soft mass-squared term allowed for the inflaton.
All of these contributions play a role in realizing $n_s$ values in the range 0.96--0.97 preferred by WMAP.
The U(1) $R$-symmetry plays an important role in flipped SU(5) by eliminating
the troublesome dimension five proton decay. The proton decays into $e^+ \pi^0$ via dimension six operators arising from the exchange of superheavy gauge bosons with a lifetime of order $10^{34}$--$10^{36}$~years.
}

\vspace{1cm}

\section*{Introduction}

The past year has seen the successful launch of two exciting experimental efforts that promise to usher in a new era of high energy physics.  On the ground, the Large Hadron Collider (LHC) at CERN will soon be scouring the TeV energy range, diligently seeking to confirm or deny a vast number of predictions.  In the sky, the Planck Surveyor has taken up residence at the L2 Lagrange point, and is actively taking precise measurements of the Cosmic Microwave Background (CMB) to further elucidate the first moments of the universe's existence.  In high energy theory, establishing a robust and testable link between cosmology and mainstream particle physics remains an ongoing challenge.  Models of supersymmetric (SUSY) hybrid inflation \cite{Dvali:1994ms,Copeland:1994vg,Linde:1993cn} involving low-scale ($\sim$~TeV) SUSY and minimal ($N=1$) supergravity (SUGRA) \cite{Chamseddine:1982jx} are particularly attractive in this sense, providing a natural connection between these two realms \cite{Senoguz:2003zw}.  As a general feature, these models incorporate the spontaneous breaking of a SUSY gauge theory $G$ to its subgroup $H$ at the end of inflation (minimal hybrid inflation) or during inflation (shifted hybrid inflation \cite{Jeannerot:2000sv}).  In the simplest model~\cite{Dvali:1994ms}, the CMB temperature anisotropy $\delta T/T$ turns out to be on the order of $(M/M_P)^2$, where $M$ is the symmetry breaking scale of $G$.  Then, in order for self-consistency of the inflationary scenario to be preserved, $M$ is comparable to the scale of grand unification, $M_{GUT} \sim 10^{16}$~GeV, hinting that $G$ may constitute a grand unified theory (GUT).  Two examples of $G$ that are highly motivated via their successes in mainstream particle physics include $\left[ \text{SU(3)}\times \text{SU(2)} \times \text{U(1)} \right] \times \text{U(1)}_{B-L}$ and the so-called flipped SU(5) model~\cite{DeRujula:1980qc}, SU(5)$\times$U(1)$_X$; we will study the latter in great detail.

The flipped SU(5) model comprises a particularly suitable choice of $G$, due to its many attractive features as a GUT model~\cite{Kyae:2005nv}, as well as its connections to F-theory~\cite{Li:2009fq,Jiang:2008yf}.  In addition to lacking topological defects, flipped SU(5) also offers a natural resolution to the doublet-triplet splitting problem.  The presence of a U(1) $R$-symmetry suppresses the dimension five operators leading to proton decay.  Proton decay via dimension six couplings to superheavy gauge bosons can still occur, and may lead to proton lifetimes within reach of upcoming experiments.  In the context of inflation, $R$-symmetry also leads to a unique renormalizable superpotential, thus flipped SU(5) constitutes an exceedingly natural choice of $G$.

In addition to their connection with symmetry breaking, these hybrid inflationary models exhibit a number of other attractive features.  For a wide range of parameters, the inflaton field amplitude takes on values that do not drastically exceed $M_{GUT}$, in contrast with the trans-Planckian values obtained in the chaotic inflationary scenario.  As a result, supergravity corrections remain under control in these models \cite{Linde:1997sj,Senoguz:2004vu}.  In addition, it has been shown within the context of these models that one of the soft SUSY-breaking terms can play a significant role during inflation \cite{Senoguz:2004vu}.  This same term has also been seen to play an integral role in resolving the MSSM $\mu$ problem~\cite{Dvali:1997uq}.
Furthermore, it was recently shown~\cite{Rehman:2009nq} that inclusion of this soft term in the inflationary potential can lead to a reduced value of the scalar spectral index $n_s$.  SUSY hybrid inflation models that include only SUGRA and radiative corrections involve a well-known lower bound of $n_s\approx 0.985$, whereas the addition of a (negative) linear soft term can result in better agreement with the central value measured by the WMAP 5 yr analysis (WMAP5).  The WMAP5 best estimate of the spectral index is given in Ref.~\cite{Komatsu:2008hk} as $n_s=0.963^{+0.014}_{-0.015}$.  In the models treated here, however, the tensor-to-scalar ratio $r$ is exceedingly small ($\lesssim 10^{-7}$); thus we will use the estimate $n_s=0.967\pm 0.025$ inferred from Ref.~\cite{Komatsu:2008hk} for $r\sim 0$.  (Accordingly, we will refer to the value $n_s\approx 0.967$ as the `central value' of WMAP5, pertaining to models of this type.)
In this letter, we carefully investigate the role during inflation of a second soft SUSY-breaking term which modifies the mass of the inflaton field. Similarly to the study performed in Ref.~\cite{Rehman:2009nq}, we find that its inclusion in the inflationary potential gives rise to a scalar spectral index which is in enhanced agreement with the WMAP5 central value. It is important to note that this result is achieved by employing the canonical (minimal) K\"ahler potential, the use of which also leads to the amelioration of the so-called `$\eta$ problem.'  For other approaches to reducing the spectral index, including use of a non-minimal K\"ahler potential, see Refs.~\cite{BasteroGil:2006cm,urRehman:2006hu,Lazarides:2007dg}.  (For a recent study leading to the reduction of the spectral index in non-SUSY hybrid inflation, see Ref.~\cite{Rehman:2009wv}.)

\section*{Minimal SUSY Hybrid Inflation}

SUSY hybrid inflation models are described by the superpotential \cite{Dvali:1994ms,Copeland:1994vg}
\begin{equation} 
W=\kappa S(\Phi \overline{\Phi }-M^{2})\,,
\label{superpot}
\end{equation}
where $S$ is a gauge singlet superfield acting as the inflaton, and $\Phi$, $\overline{\Phi }$ are conjugate superfields transforming nontrivially under some gauge group $G$, and provide the vacuum energy associated with inflation.  If one assumes an additional U(1)$_R$ `$R$-symmetry,' it can readily be seen that $W$ in Eq.~(\ref{superpot}) is the unique superpotential involving these superfields that is invariant under both U(1)$_R$ and $G$ at the renormalizable level.

The canonical K\"ahler potential may be written as
\begin{equation}
K=  |S|^{2}+ |\Phi|^{2} + |\overline{\Phi}|^{2}.  \label{kahler}
\end{equation}
Recent analyses have used a non-minimal K\"ahler potential 
to obtain better agreement with WMAP5 (see, for example, Refs. \cite{BasteroGil:2006cm,urRehman:2006hu}).  However, as we will see, the minimal K\"ahler potential will be sufficient for our purposes.  The SUGRA scalar potential is given by
\begin{equation}
V_{F}=e^{K/m_{P}^{2}}\left(
K_{ij}^{-1}D_{z_{i}}WD_{z^{*}_j}W^{*}-3m_{P}^{-2}\left| W\right| ^{2}\right),
\label{VF}
\end{equation}
where $z_{i}$ represent the bosonic components of the superfields, $z%
_{i}\in \{\phi , \overline{\phi }, s,\cdots\}$, and where we have defined
\begin{equation*}
D_{z_{i}}W \equiv \frac{\partial W}{\partial z_{i}}+m_{P}^{-2}\frac{\partial K}{\partial z_{i}}W, \,\,\,\,\,\, K_{ij} \equiv \frac{\partial ^{2}K}{\partial z_{i}\partial z_{j}^{*}},
\end{equation*}
$D_{z_{i}^{*}}W^{*}=\left( D_{z_{i}}W\right) ^{*}$ and $m_{P}=M_P/\sqrt{8\pi} \simeq 2.4\times 10^{18}$~GeV is the reduced Planck mass.  In the D-flat direction, $|\phi|=|\overline{\phi}|$; then, using Eqs.~(\ref{superpot})--(\ref{VF}), we may write the tree level global SUSY potential as
\begin{equation}
V_{F}= \kappa^2\,(M^2 - \vert \phi\vert^2)^2 + 2\kappa^2 \vert s \vert^2 \vert \phi \vert^2 .
\label{VF2}
\end{equation}
A suitable choice of initial conditions ensures that the fields become trapped in the (flat) valley along $|s|>s_c=M$, $|\phi|=|\overline{\phi}|=0$.  The gauge group $G$ is unbroken along this valley; however, only the constant term $V_0=\kappa^2 M^4$ is present at tree level, thus SUSY is broken during inflation.  Taking into account leading order SUGRA corrections, as well as radiative corrections \cite{Dvali:1994ms} and soft SUSY-breaking terms, the flat valley is lifted and the inflaton $s$ rolls toward smaller values.  Upon reaching its critical value $s_c=M$, inflation ends via waterfall as the fields rapidly transition to the global SUSY minimum located at $|s|=0$, $|\phi|=|\overline{\phi}|=M$, beginning damped oscillations about this vacuum to reheat the universe.

By including the various correction terms, we may write the scalar potential along the inflationary trajectory (i.e. $|\phi|=|\overline{\phi}|=0$) as
\begin{equation}
V \simeq \kappa ^{2}M^{4}\left( 1 + \left( \frac{M}{m_{P}}\right) ^{4}\frac{x^{4}}{2}+\frac{\kappa ^{2}\mathcal{N}}{8\pi ^{2}}F(x) + a\left(\frac{m_{3/2}\,x}{\kappa\,M}\right) + \left( \frac{M_S\,x}{\kappa\,M}\right)^2\right) ,
\label{scalarpot}
\end{equation}
where 
\begin{equation}
F(x)=\frac{1}{4}\left( \left( x^{4}+1\right) \ln \frac{\left( x^{4}-1\right)}{x^{4}}+2x^{2}\ln \frac{x^{2}+1}{x^{2}-1}+2\ln \frac{\kappa ^{2}M^{2}x^{2}}{Q^{2}}-3\right)
\end{equation}
represents radiative corrections, and 
\begin{equation}
a = 2\left| 2-A\right| \cos [\arg s+\arg (2-A)].
\label{a}
\end{equation}
Here, $x\equiv |s|/M$ parametrizes the inflaton field, $\mathcal{N}$ is the dimensionality of the representation of fields $\phi$ and $\overline{\phi}$ with respect to $G$, $m_{3/2}\simeq 1$~TeV is the gravitino mass, and $Q$ is the renormalization scale.  The last two terms in Eq.~(\ref{scalarpot}) arise as soft SUSY-breaking linear and mass-squared terms, respectively, and their form here is derived from a gravity-mediated SUSY-breaking scheme~\cite{Dvali:1997uq}.  The coupling $A-2$ is the coefficient of the linear soft term in the Lagrangian~\cite{Nilles:1983ge}, and $A\sim O(1)$ is expected.
  As can be seen from Eq.~(\ref{a}), the dimensionless parameter $a$ can have any sign; indeed, in a recent analysis \cite{Rehman:2009nq}, we have pointed out that choosing the negative sign can result in good agreement with the central value of the spectral index $n_s$ as measured by WMAP5 \cite{Komatsu:2008hk}.  In the case of TeV-scale soft masses, this holds as long as $a<0$ with a magnitude $|a|\gtrsim 10^{-6}$.  In our calculations, we will consider three representative examples, $a=0$, $\pm 1$.  Note that a constant value of $a$ corresponds to negligible variation of $\arg s$, which can be achieved by employing an appropriate choice of initial condition \cite{urRehman:2006hu}.

In this letter, we investigate the effect of a negative soft mass-squared term for the inflaton (i.e. $M_S^2<0$) occurring at intermediate scales.  The physical inflaton mass is given by
\begin{equation}
m_\text{inf}=\sqrt{2(\kappa^2 M^2+M_S^2)},
\end{equation}
and so it is necessary that $\sqrt{2}|M_S|$ be significantly smaller than the tree-level mass $\sqrt{2}\kappa M$ in order to ensure that $m_\text{inf}$ remains real.  In our numerical calculations, this condition is met throughout the range of interest without needing to impose the restriction by hand.  Tachyonic soft masses have previously been utilized to reduce fine tuning in the MSSM~\cite{Feng:2005ba}.  
As we will see, a negative soft mass-squared term can produce a reduction of the spectral index similar to the case of $a<0$, $M_S^2>0$ treated in Ref.~\cite{Rehman:2009nq}, for a range of intermediate mass scales depending in part upon the sign of $a$.  The soft mass scales employed here are reminiscent in magnitude of split SUSY models, in which the scalar superpartners attain masses at a separate (typically much larger) scale compared to their fermionic counterparts~\cite{ArkaniHamed:2004fb}.  A crucial difference between our models and split SUSY is the sign of the soft mass-squared term, which is positive in split SUSY.

The number of e-foldings after a given comoving scale $l$ has crossed the horizon is given by
\begin{equation}
N_{l}=2\left( \frac{M}{m_{P}}\right) ^{2}\int_{x_e}^{x_{l}}\left( \frac{V}{\partial _{x}V}\right) dx , \label{Nl}
\end{equation}
where $x_l$ and $x_e$ denote the values of $x$ at $l$ and at the end of inflation, respectively.  Inflation may end via a waterfall induced at $x_c\equiv |s_c|/M=1$ if the slow-roll approximation holds.  However, the slow-roll conditions are typically violated at $x$ slightly larger than unity, in which case inflation ends via slow-roll breakdown somewhat before the waterfall would occur.  The number of e-folds after the scale $l_0=2\pi/k_0$ corresponding to the WMAP pivot scale $k_0=0.002$~Mpc$^{-1}$ exits the horizon is approximately given by
\begin{equation}
N_{0} \simeq 53+\frac{1}{3}\ln \left( \frac{T_{r}}{10^{9}~{\rm GeV}}\right) +\frac{2}{3}\ln \left( \frac{V(x_0)^{1/4}}{10^{15}~{\rm GeV}}\right),   \label{N0}
\end{equation}
where we have assumed a matter-dominated reheating phase.  The reheat temperature $T_r$ depends on the mechanism by which the universe was reheated after the end of inflation.  As we do not wish to assume a specific reheating mechanism, we will use a constant value $T_r\simeq 10^9$~GeV, although lower values of $T_r$ may also be considered.

In Eq.~(\ref{N0}) and throughout, a subscript `0' denotes quantities taken at the pivot scale $k_0$.  In particular, the amplitude of the primordial curvature perturbation is given by
\begin{equation}
\Delta_{\mathcal{R}}=\frac{M}{\sqrt{6}\,\pi \,m_{P}^{3}}\left( \frac{V^{3/2}}{|\partial
_{x_{0}}V|}\right) ,  \label{curv}
\end{equation}
and has been measured by WMAP5 to be $\Delta_{\mathcal{R}}=4.91\times 10^{-5}$ at $k_0$ \cite{Komatsu:2008hk}.  This important constraint is useful in the determination of the value $x_0$ at the start of observable inflation.

In the case of $a\geq 0, M_S^2 \geq 0$, the potential in Eq.~(\ref{scalarpot}) monotonically increases with $x$.  In contrast, if either (or both) of these parameters is negative, a metastable (SUSY-broken) vacuum may be induced in some region of parameter space.  However, it is possible to obtain enough e-foldings for successful inflation even if the local potential maximum occurs very close to $x_c=1$.  In our calculations, we will assume that the inflaton has successfully escaped any metastable vacua, and the solutions presented will correspond to the last 50--60 or so e-foldings occurring in a region where the potential monotonically increases with $x$.  When a metastable vacuum is present in the potential, inflation begins near the local maximum where the potential is concave downward, constituting `hilltop' inflation \cite{Boubekeur:2005zm}.  This behavior is also exhibited in many cases in which the potential possesses points of inflection but lacks additional extrema.

The usual slow-roll parameters are given by
\begin{equation}
\epsilon =\frac{m_{P}^{2}}{4\,M^2}\left( \frac{\partial_x V }{V}\right) ^{2},\,\,
\eta =\frac{m_{P}^{2}}{2\,M^2}\left( \frac{\partial_x^2 V }{V}\right).\label{slowroll}
\end{equation}
Within the framework of the slow-roll approximation (i.e. $\epsilon,|\eta|\ll 1$), the scalar spectral index appears as
\begin{equation}
n_{s} \simeq 1 -6\epsilon +2\eta \,\, \simeq \,\, 1 + 6\left( \frac{M}{m_{P}}\right)
^{2}x_{0}^{2}+\left( \frac{m_{P}}{M}\right) ^{2}\left[ \frac{\kappa ^{2}%
\mathcal{N}}{8\pi ^{2}} \partial _{x_0}^{2}F(x_0)+ 2 \left(\frac{M_S}{\kappa\,M}\right)^2 \right]\,\label{ns}
\end{equation}
to leading order in the slow-roll parameters.  (In our numerical calculations, we have used the next-to-leading order expressions in the slow-roll parameters for $n_s$ and other quantities, for additional precision.)  The slow-roll parameter $\epsilon$ is quite small in these models, and is consistently subdominant in comparison to $\eta$.  Thus we have suppressed the contribution of the $\epsilon$ term in Eq.~(\ref{ns}).  The terms arising from radiative corrections and the soft mass-squared term both yield a negative contribution to $(n_s-1)$, encouraging a red-tilted spectrum.  (Notice that $M_S^2>0$ leads to an enhancement rather than reduction of the spectral index, and such a term pushes $n_s$ farther from the WMAP5 central value.)  In addition, a subtle contribution from the soft linear term can play a significant role in determining the behavior of $n_s$, arising indirectly via the WMAP5 normalization of $\Delta_{\mathcal{R}}$ and the shift of $x_0$ with a change of $a$ \cite{Rehman:2009nq}.

\section*{Flipped SU(5)}

There are a number of natural choices for the gauge group $G$.  As mentioned previously, $G$ can be associated with a U(1)$_{B-L}$ symmetry, a case which is motivated in terms of generating the observed baryon asymmetry via leptogenesis.  We may alternatively identify $G$ with a GUT, such as SO(10) or SU(5)\ \cite{Kyae:2004ft}.  An attractive choice is the so-called flipped SU(5) model, SU(5)$\times$U(1)$_X$, which possesses a number of advantages over other models.  In this case, symmetry breaking is induced by a 10-plet, i.e. $\Phi={\bf 10}_H$ and $\mathcal{N}=10$, and the superpotential is given by~\cite{Kyae:2005nv}
\begin{eqnarray} \label{Wfull}
W &=& \kappa S\left[{\bf 10}_H{\bf \overline{10}}_H-M^2\right] \nonumber \\
&+& \lambda_1{\bf 10}_H{\bf 10}_H{\bf 5}_h+\lambda_2\overline{{\bf 10}}_H\overline{{\bf 10}}_H\overline{{\bf 5}}_h \\
&+& y_{ij}^{(d)}{\bf 10}_i{\bf 10}_j{\bf 5}_h +y_{ij}^{(u,\nu)}{\bf 10}_i\overline{{\bf 5}}_j\overline{{\bf 5}}_h + y_{ij}^{(e)}{\bf 1}_i\overline{{\bf 5}}_j{\bf 5}_h \nonumber ~,
\end{eqnarray}
where $y^{(d)},y^{(u,\nu)},y^{(u)}$ denote the Yukawa couplings for quarks and charged leptons.  (For a discussion of neutrino masses in these models, see Ref.~\cite{Kyae:2005nv}.)
The first line in Eq.~(\ref{Wfull}) is relevant for inflation, and corresponds to the superpotential in Eq.~(\ref{superpot}).  The terms in the second line are involved in the solution of the doublet-triplet splitting problem, and the last line contains terms that generate masses for quarks and charged leptons.  Following~\cite{Kyae:2005nv}, the charges of the multiplets with respect to the U(1)$_R$ $R$-symmetry are assigned as
\begin{equation}
(S,{\bf 10}_H,{\bf \overline{10}}_H,{\bf 5}_h,{\bf \overline{5}}_h,{\bf 10}_i,{\bf \overline{5}}_i,{\bf 1}_i)=(1,0,0,1,1,0,0,0).
\end{equation}
Note that U(1)$_R$ forbids the appearance of the bilinear term proportional to ${\bf 5}_h {\bf \overline{5}}_h$ in the superpotential~\cite{Kyae:2005nv}.  The color Higgs triplets in ${\bf 10}_H$ and ${\bf \overline{10}}_H$ become superheavy due to the couplings $\lambda_1{\bf 10}_H{\bf 10}_H{\bf 5}_h$ and $\lambda_2\overline{{\bf 10}}_H\overline{{\bf 10}}_H\overline{{\bf 5}}_h$, while the MSSM Higgs doublets remain light.
Furthermore, the absence of the ${\bf 5}_h {\bf \overline{5}}_h$ term in Eq.~(\ref{Wfull}) leads to the dimension-five proton decay amplitude being heavily suppressed (by an additional factor $\mu / M_{GUT}$, where $\mu \sim$~TeV is the MSSM $\mu$ term).  The $\mu$ problem can be solved, as previously discussed in Ref.~\cite{Kyae:2005nv}, by exploiting the Giudice-Masiero mechanism~\cite{Giudice:1988yz}.  Proton decay proceeds via the dimension-six operator, ($p\rightarrow \pi^0 e^+$)\footnote{For a recent discussion of proton decay in the context of non-SUSY inflation models, see Ref.~\cite{Rehman:2008qs}.}.  (For further discussion of proton decay in the context of flipped SU(5), see Ref.~\cite{Shafi:2006dm}.)  Interestingly, it is the same U(1)$_R$ symmetry that we have already seen to be important for SUSY hybrid inflation that also alleviates some of the typical GUT problems.

The flipped SU(5) model exhibits yet more desirable features.  In contrast to the U(1)$_{B-L}$ model, flipped SU(5) benefits from the absence of topological defects ($B-L$ strings as well as monopoles).  As opposed to the SO(10) and standard SU(5) models, only the minimal Higgs sector is needed in flipped SU(5).  Finally, flipped SU(5) models have also been seen to have intimate connections to F-theory~\cite{Li:2009fq,Jiang:2008yf}.

\section*{Results and Discussion}

\begin{figure}[!p]
\begin{tabular}{cc}
\includegraphics[width=7.5cm]{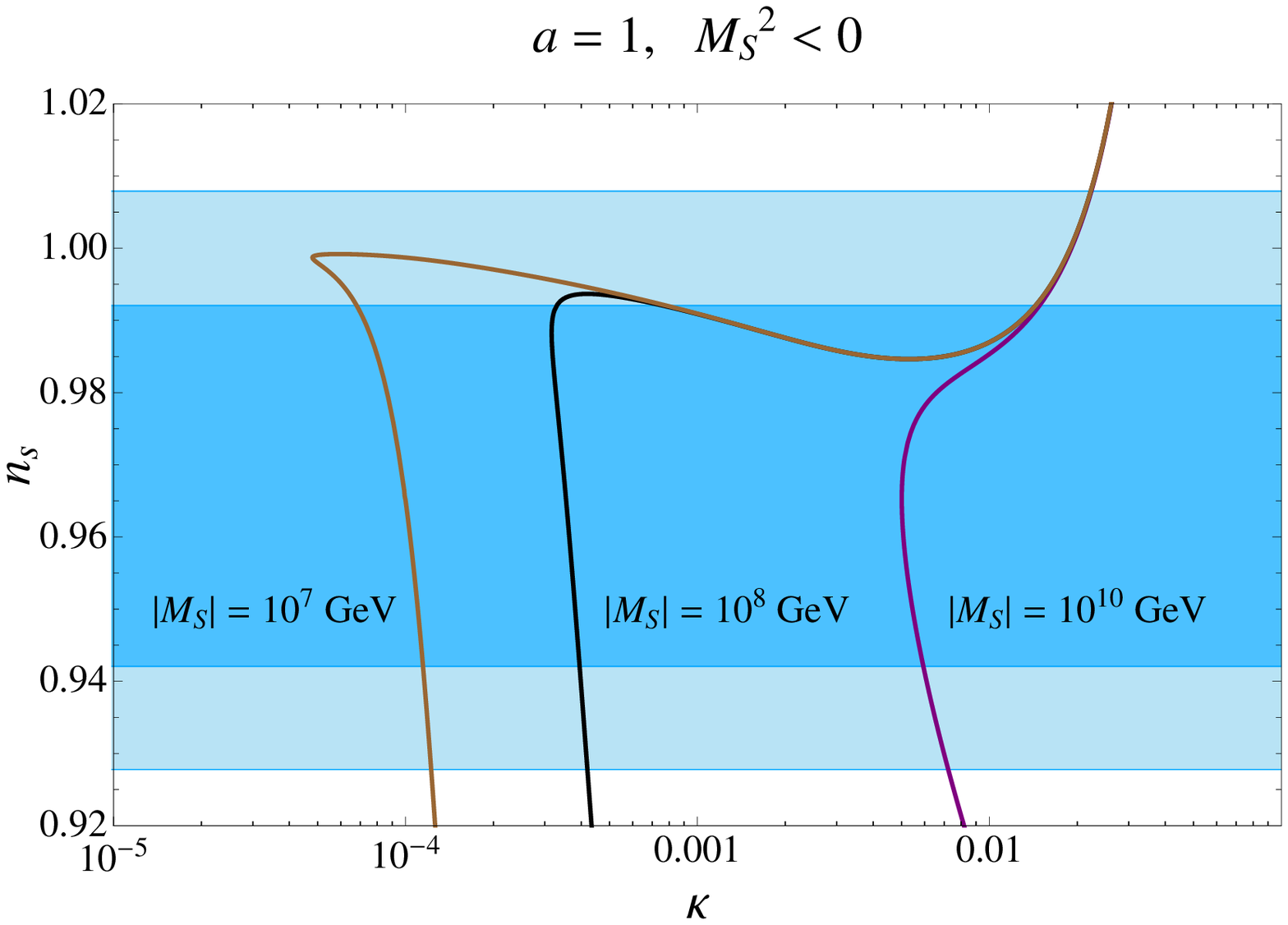} & \includegraphics[width=7.5cm]{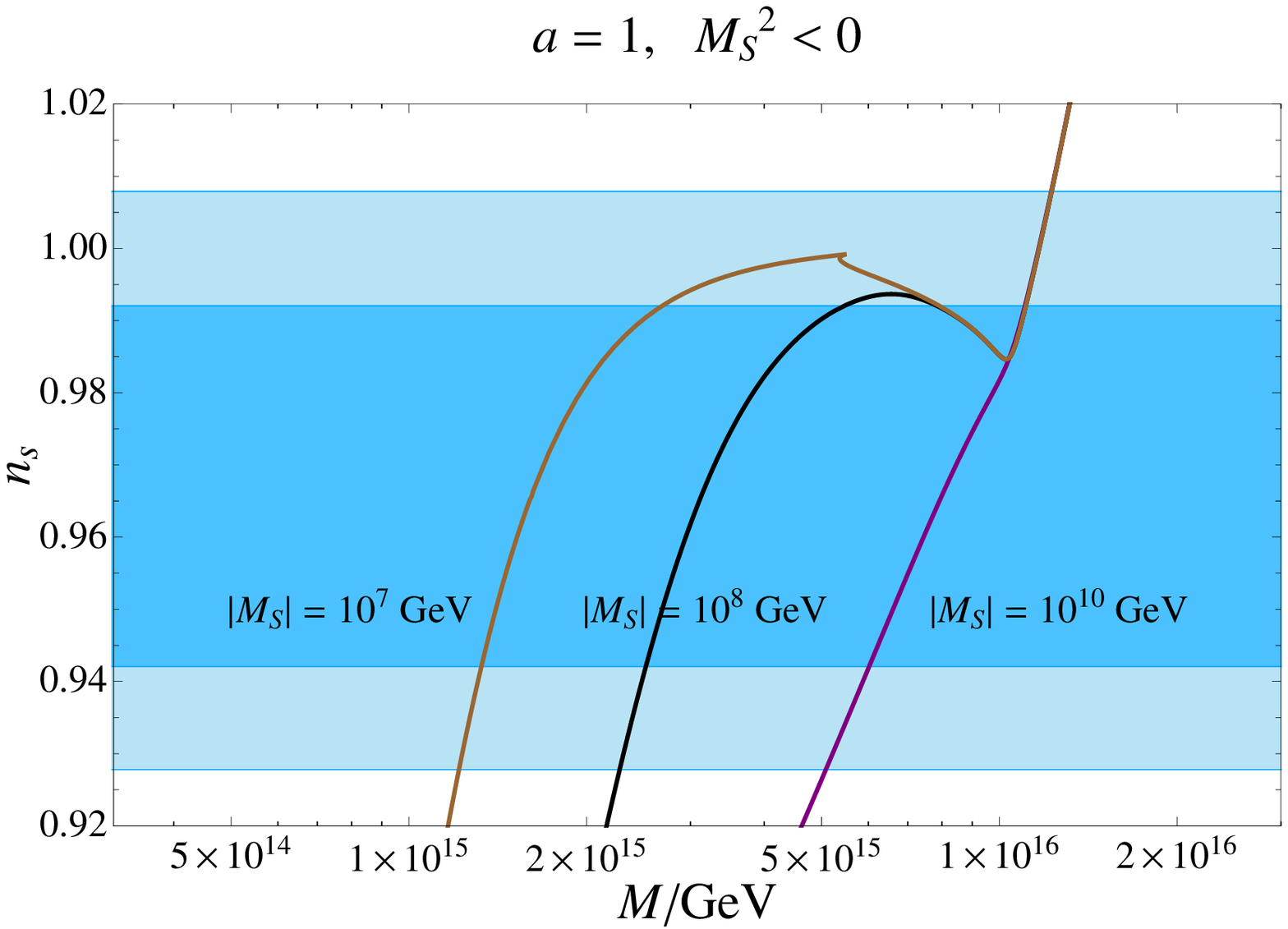} \\
(a) & (b) \\
\end{tabular}
\caption{$n_s$ vs. $\log_{10}\kappa$ and $\log_{10}(M/\text{GeV})$, for $|M_S| = 10^{10},10^{8},10^{7}$~GeV (purple, black, and brown curves, respectively) and $a = 1$, with $a$ given by Eq.~(\ref{a}).  These panels include the WMAP5 68\% and 95\% confidence level contours for the value of the spectral index $n_s$ \cite{Komatsu:2008hk}.} \label{rnsk_p}
\end{figure}

\begin{figure}[b]
\begin{tabular}{cc}
\includegraphics[width=7.5cm]{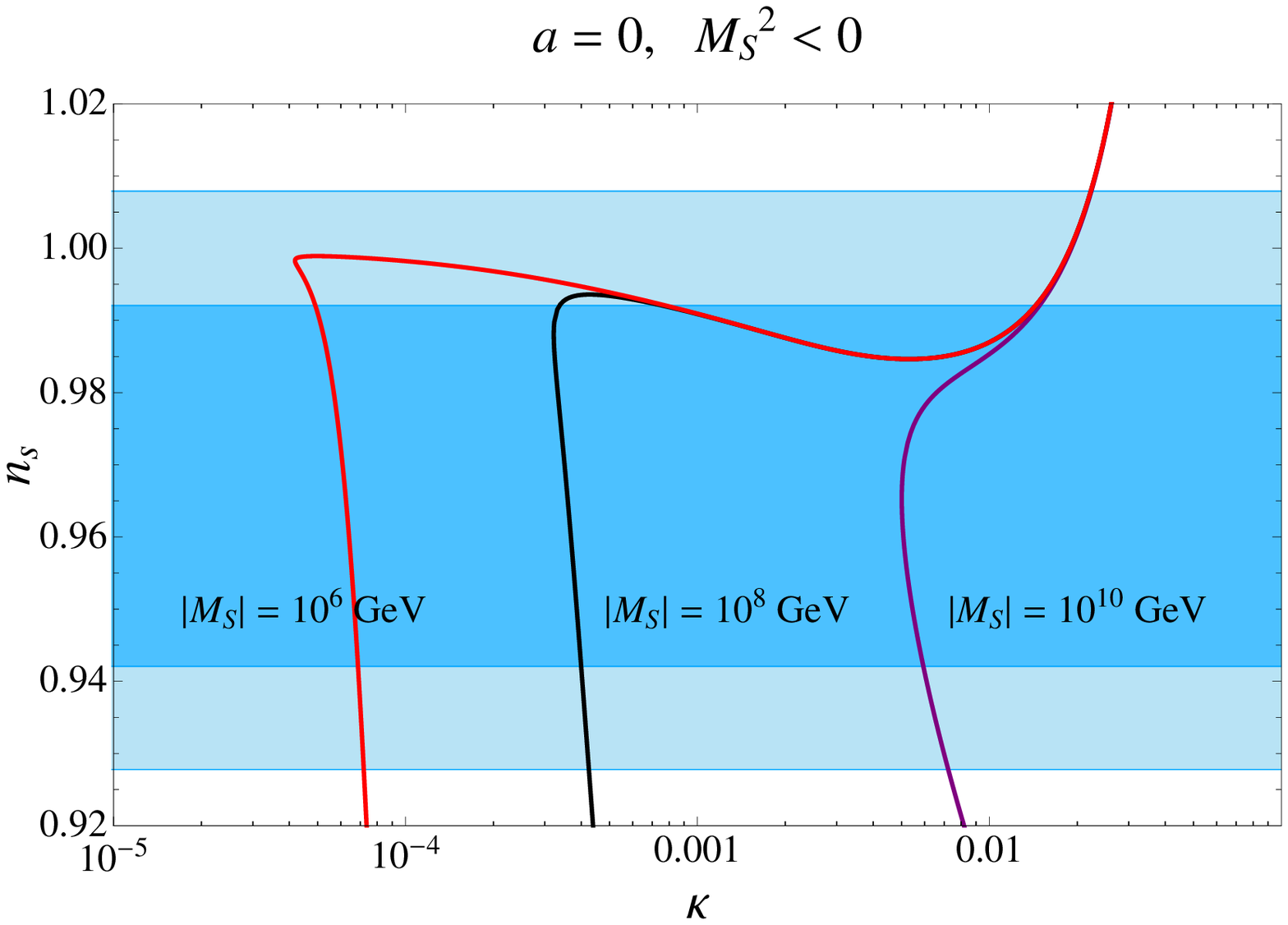} & \includegraphics[width=7.5cm]{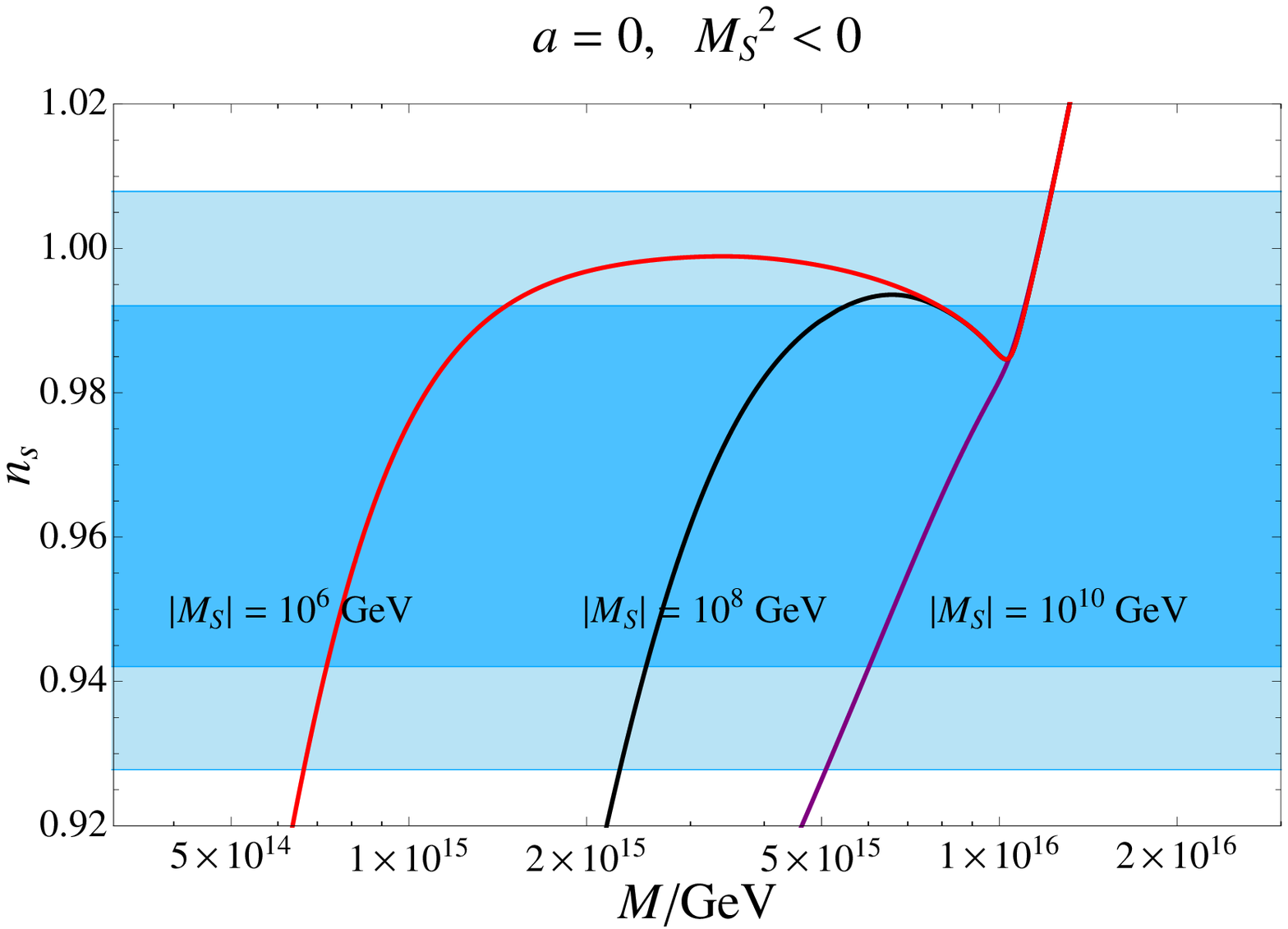} \\
(a) & (b) \\
\end{tabular}
\caption{$n_s$ vs. $\log_{10}\kappa$ and $\log_{10}(M/\text{GeV})$, for $|M_S| = 10^{10},10^{8},10^{6}$~GeV (purple, black, and red curves, respectively) and $a = 0$, with $a$ given by Eq.~(\ref{a}).  These panels include the WMAP5 68\% and 95\% confidence level contours for the value of the spectral index $n_s$ \cite{Komatsu:2008hk}.} \label{rnsk_z}
\end{figure}

\begin{figure}[t]
\begin{tabular}{cc}
\includegraphics[width=7.5cm]{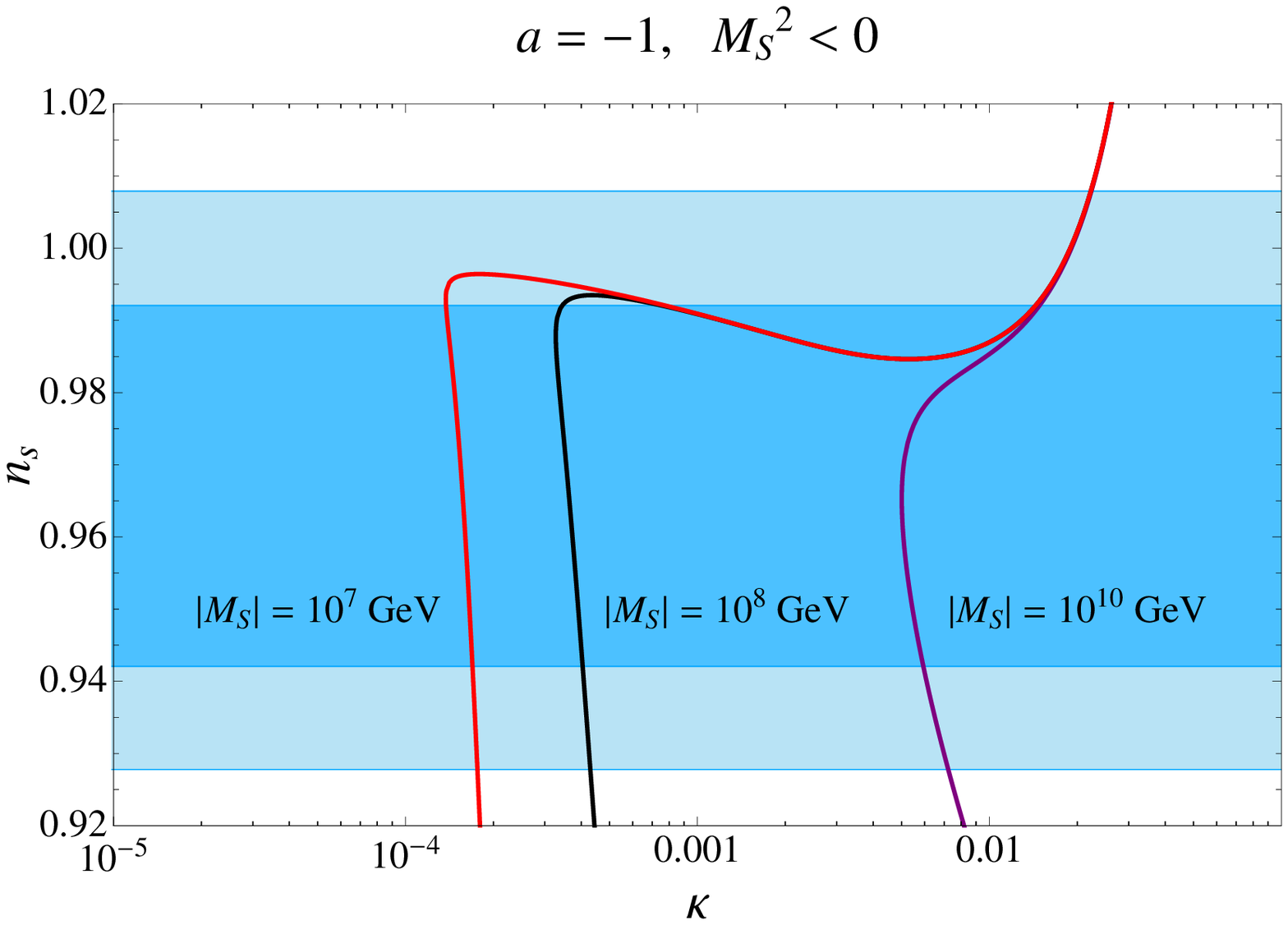} & \includegraphics[width=7.5cm]{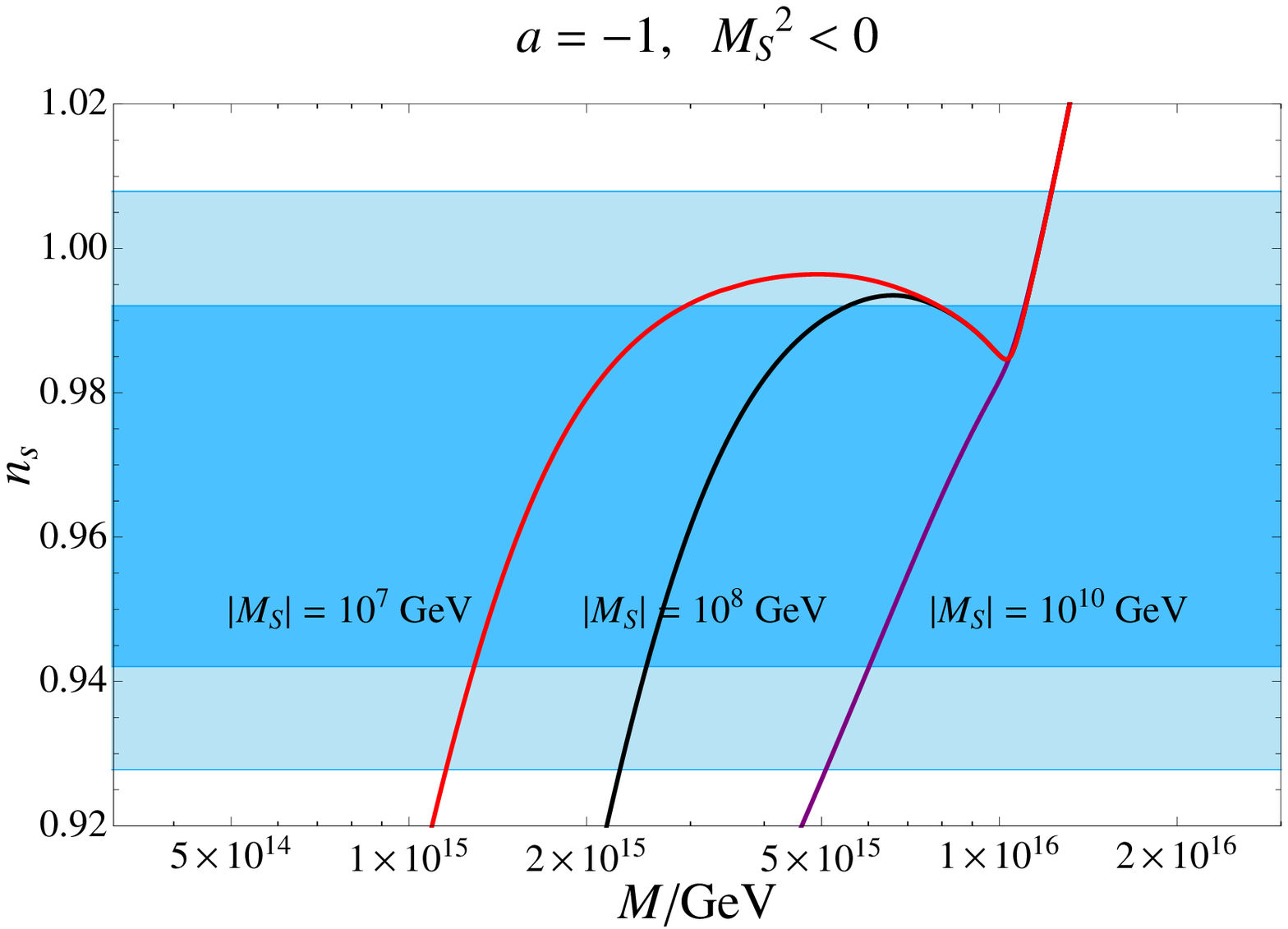} \\
(a) & (b) \\
\end{tabular}
\caption{$n_s$ vs. $\log_{10}\kappa$ and $\log_{10}(M/\text{GeV})$, for $|M_S| = 10^{10},10^{8},10^{6}$~GeV (purple, black, and red curves, respectively) and $a = -1$, with $a$ given by Eq.~(\ref{a}).  The red curve represents the limiting case; no substantial change is exhibited beyond $|M_S| = 10^{6}$~GeV for $a=-1$.  These panels include the WMAP5 68\% and 95\% confidence level contours for the value of the spectral index $n_s$ \cite{Komatsu:2008hk}.} \label{rnsk_n}
\end{figure}

Using Eqs.~(\ref{scalarpot})--(\ref{slowroll}), we have performed a series of numerical calculations to determine $M$ as a function of $\kappa$ over a range of $|M_S|$ for the representative cases $a=0$,~$\pm 1$ and $\mathcal{N}=10$.  Subsequently, we have calculated various other quantities (e.g. $n_s$) that depend on these parameters.  The results of these calculations are presented in Figs.~\ref{rnsk_p}--\ref{rnsk_n}.  We find that a red-tilted spectrum in good agreement with the WMAP5 central value, $n_s\simeq 0.967$, is obtained for $|M_S|\lesssim 10^{11}$~GeV in all cases of $a$.  For $|M_S|\gtrsim 10^{11}$~GeV, the coupling $\kappa$ begins to take on large values ($\gtrsim 10^{-1}$), supergravity corrections become increasingly important, and the spectral index $n_s$ approaches unity.  The lowest value of $|M_S|$ leading to $n_s<1$ depends heavily upon $a$.

\begin{figure}[b]
\begin{tabular}{cc}
\includegraphics[width=7.5cm]{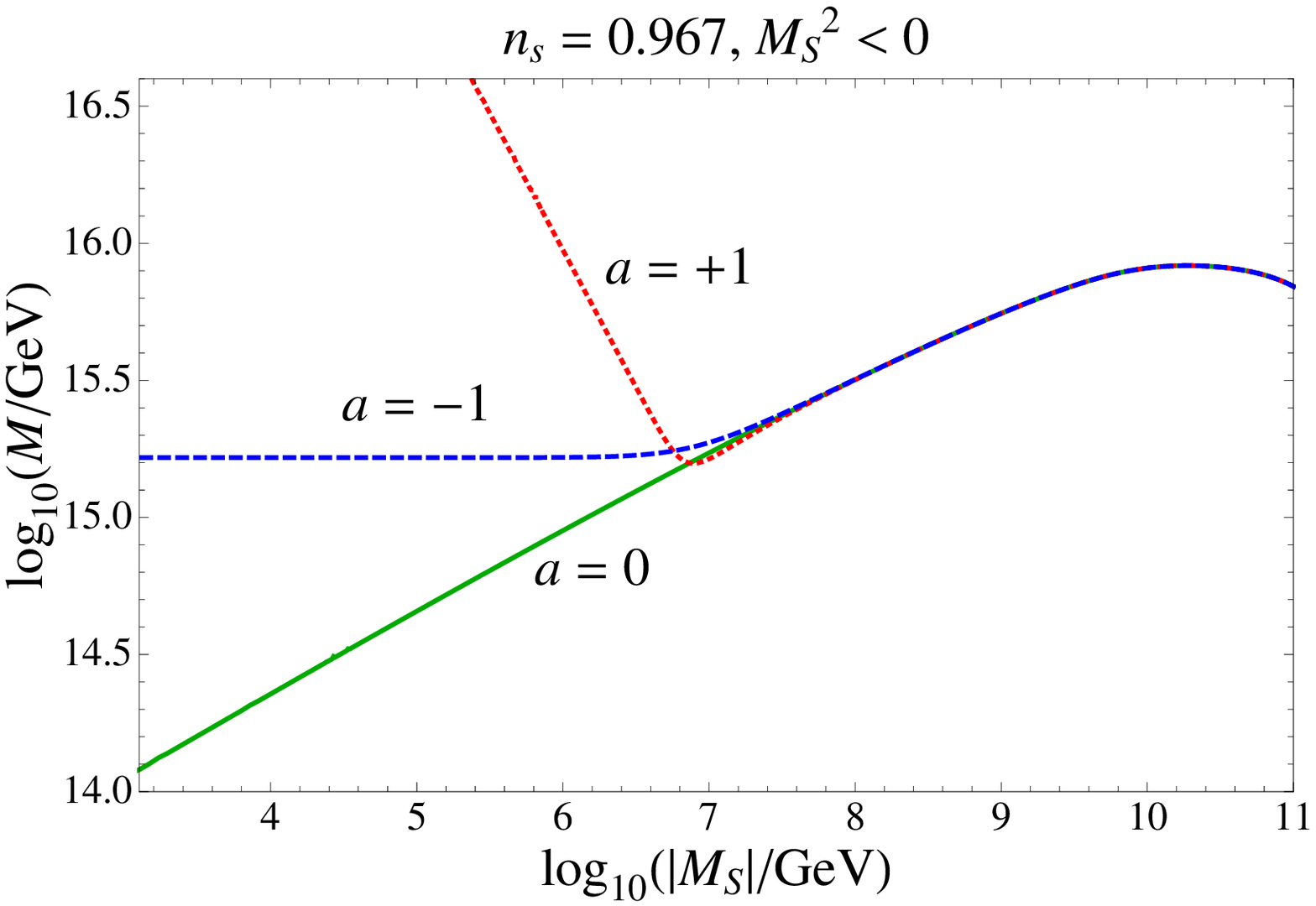} & \includegraphics[width=7.5cm]{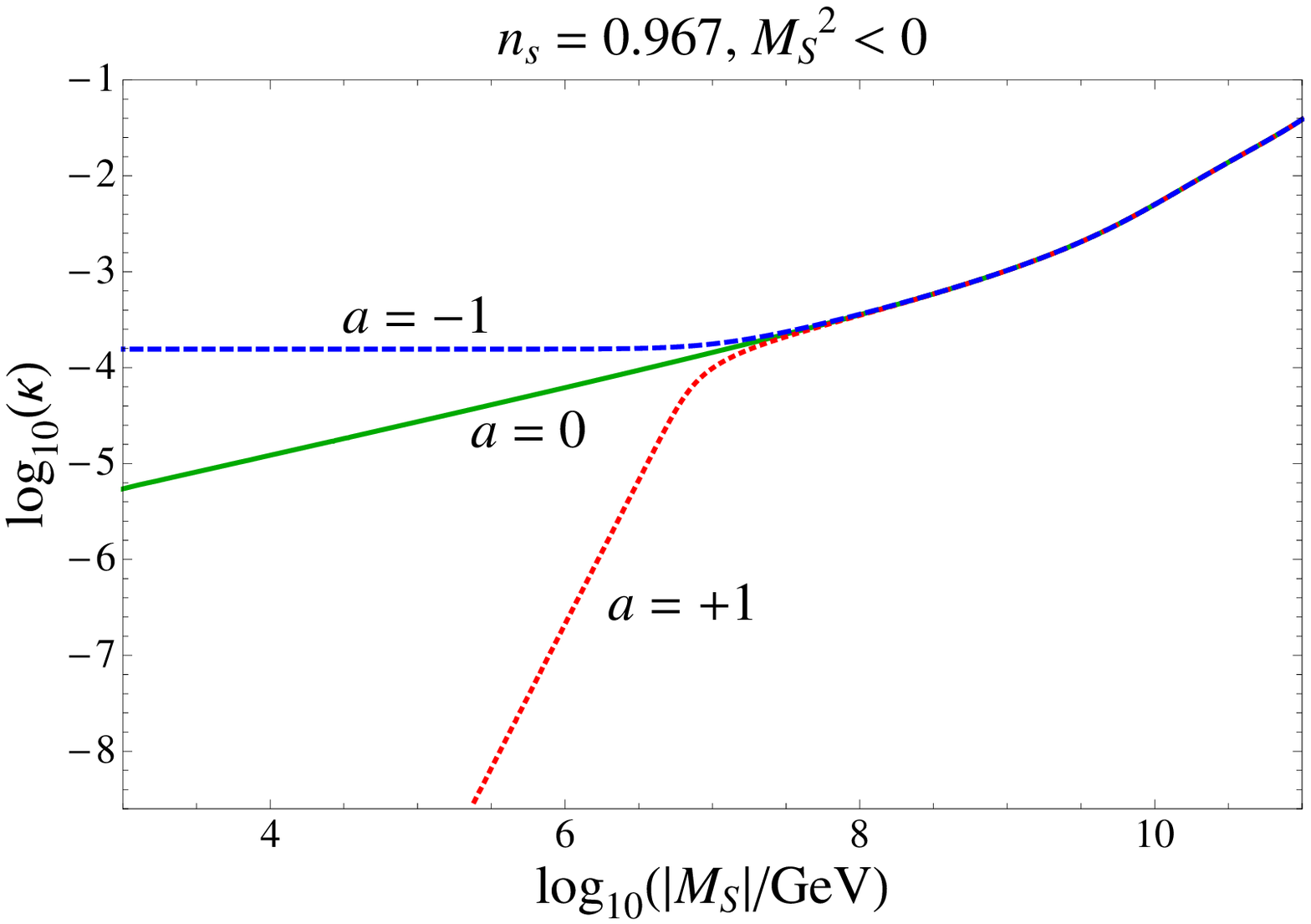} \\
(a) & (b) \\
\end{tabular}
\caption{$\log_{10}(M/\text{GeV})$ and $\log_{10}\kappa$ vs. $\log_{10}(|M_S|/\text{GeV})$, with $n_s$ fixed at the central value 0.967 for $r\sim 0$ \cite{Komatsu:2008hk}.} \label{MkMS}
\end{figure}

The left (right) panels of Figs.~\ref{rnsk_p}, \ref{rnsk_z} and \ref{rnsk_n} show the behavior of $n_s$ with respect to $\kappa$ ($M$) for $a=1$, 0 and $-1$, respectively.  In all of the cases depicted, a red tilted spectral index can be achieved, spanning the WMAP5 $2\sigma$ range for all $|M_S|$ values shown.  
As $|M_S|$ decreases, a clear trend toward smaller values of both $\kappa$ and $M$ is exhibited.  For the highest values of $|M_S|$ (say $10^{10}$~GeV), the $a$-term is subdominant and all cases of $a$ are degenerate.  This term begins to yield a sizable contribution around $|M_S|\sim 10^{7}$~GeV and the predictions for $a=0,\pm 1$ begin to diverge, as seen in Fig.~\ref{MkMS}.  In these panels, $n_s$ is held fixed at the central value 0.967 for ease in exploring an extended region of parameter space.  For $a=+1$, we find that a red-tilted spectrum is possible for $|M_S|\gtrsim 10^{5}$~GeV.  For $a=0$, good agreement with WMAP5 can be obtained for $|M_S|$ values as low as the TeV scale.  In the case of $a=-1$, the WMAP5 central value of $n_s$ can be obtained with arbitrarily small $|M_S|$ if we maintain that $M_S^2<0$.  As we have already mentioned, a suitably red-tilted spectrum can be obtained with $a=-1$ even in the case of a slightly positive soft mass-squared term (i.e. TeV scale) \cite{Rehman:2009nq}, and so this result is expected.  Indeed, for $|M_S|\lesssim 10^6$~GeV, the $a$-term begins to dominate over the mass-squared term, and no further significant change in behavior is exhibited.  (This result was also seen in the full calculations where $n_s$ is free to vary.)

To understand the behavior exhibited in Fig.~\ref{MkMS}, it is useful to analytically examine some approximate formulae.  For clarity, we will fix the spectral index at the central value $n_s\approx 0.967$ for the purposes of this discussion.  In the slow roll limit, the potential is dominated by the vacuum term, $V\simeq \kappa^2 M^4$, and Eq.~(\ref{curv}) becomes
\begin{equation}
\Delta_{\mathcal{R}} \simeq \frac{\kappa^2}{2\sqrt{6}\pi} \left( \frac{M}{m_P} \right)^4 \left[ \kappa\left(\frac{M}{m_P}\right)^5\,x_0^3 + \frac{\kappa^3 \mathcal{N}\,\partial_{x_0}F(x_0)}{16\pi^2} \left(\frac{M}{m_P}\right) + \frac{a\,m_{3/2}}{2 m_P} + \frac{M_{S}^2\,x_0}{\kappa M m_P} \right]^{-1}.
\end{equation}
For $a = 0$, the loop correction and soft mass-squared terms remain important for most $M_S$ 
values, until the SUGRA correction begins to dominate (around $M_S \sim 10^{11}$~GeV) which drives the curve downward in Fig.~\ref{MkMS}(a). Taking the soft mass-squared term to be comparable to 
the loop correction term in $\partial_{x_0}V(x_0)$ 
and using the expression for $n_s$ in Eq.~(\ref{ns}), we arrive at the following result
\begin{eqnarray}
|M_{S}| &\simeq& \frac{(1-n_s)\,\sqrt{\pi^2\,\ln[4]/\mathcal{N}}}{\ln[4]+|\partial_{x_0}^2 F(x_0)|}
\,\left( \frac{M}{m_P} \right)^2\,M, \\
|M_{S}| &\simeq& \frac{\kappa^3\,\mathcal{N} \,\sqrt{\ln[2]\,(\ln[4]+|\partial_{x_0}^2 F(x_0)|)}}
{8\,\pi^2\,(1-n_s)^{1/2}}\,m_P,
\end{eqnarray}
where $|\partial_{x_0}^2 F(x_0)|$ varies from 5 to 17 for $M_S/\text{GeV} \sim 10^{7}$--$10^3$ in order to
obtain enough e-foldings.

As we have already noted, the curves corresponding to $|a| \sim 1$ begin to diverge from the $a=0$ result at a scale around $M_S \sim 10^{7}$~GeV, below which the linear soft SUSY-breaking term is important.  The $a = +1$ case is particularly interesting, as it generates
$M$ values in a range which is important for proton decay considerations. In this case, the loop correction becomes supressed due to small
$\kappa$ values, as shown in Fig.~\ref{MkMS}(b); this greatly simplifies the analytical solution. Here we take the soft mass-squared term to be comparable to 
linear soft SUSY-breaking term in $\partial_{x_0}V(x_0)$:
\begin{equation}
\frac{a\,m_{3/2}}{2 m_P} \simeq \frac{|M_{S}|^2}{\kappa M m_P},  \,\,\,  
n_{s} \simeq 1 - 2\,\left( \frac{m_{P}}{M}\right) ^{2} \, \left(\frac{|M_{S}|}{\kappa\,M}\right)^2
+ 6\left( \frac{M}{m_{P}}\right)^2.
\end{equation}
It turns out that the two soft terms cancel one another up to a few percent, and then their sum partially cancels with the SUGRA term to generate the measured curvature perturbation 
$\Delta_{\mathcal{R}} = 4.91\times 10^{-5}$. We then obtain the following relations for $M_S$ in terms of $M$ and $\kappa$:
\begin{eqnarray}
|M_S| &\simeq& \left( \frac{2}{1-n_s} \right)^{1/2}\,\left(\frac{a\,m_{3/2}}{2\,M}\right)\,m_P, \\
|M_S| &\simeq& \left( \frac{2}{1-n_s} \right)^{1/6}\,\left(\frac{a\,m_{3/2}}{2}\right)^{2/3}
\,(\kappa\,m_P)^{1/3}.
\end{eqnarray}

For the $a = -1$ case, the SUGRA and soft mass-squared terms become negligible below $M_S \sim 10^7$~GeV, and the behavior of the curve decouples from $|M_S|$.
Taking the loop term to be comparable to the soft linear term, we obtain asymptotic $M$ and $\kappa$ values given by
\begin{eqnarray}
M &\simeq&  \left( \frac{|a|\,m_{3/2}\,m_P^3\,|\partial_{x_0}^2 F(x_0)|^{3/2}}{2^{5/2}\,(1-n_s)^{3/2}\,\pi\,\ln[2]/\sqrt{\mathcal{N}}} 
\right)^{1/4} \simeq 1.65 \times 10^{15} \text{GeV},\\
\kappa &\simeq& \left( \frac{2^{7/2}\,|a|\,m_{3/2}\,(1-n_s)^{1/2}\,\pi^3}{m_P\,\mathcal{N}^{3/2}\,\ln[2]\,|\partial_{x_0}^2 F(x_0)|^{1/2}} \right)^{1/4} \simeq 1.6 \times 10^{-4},
\end{eqnarray}
where $|\partial_{x_0}^2 F(x_0)| \simeq 5$ remains constant below $M_S \sim 10^7$ GeV.

In addition to the spectral index $n_s$, WMAP measures a variety of other quantities holding a great deal of information about the early universe.  One such parameter, the tensor-to-scalar ratio $r$, contains information about the amplitude of primordial gravitational waves, as well as the energy scale of inflation via $V(\phi_0)^{1/4}\simeq 3.3\times 10^{16} \text{ GeV} \cdot r^{1/4}$.  While current experimental bounds allow for a wide range of $r$-values, this situation promises to improve in the near future with the recent launch of the Planck Surveyor.  The SUSY hybrid models which we consider here predict exceedingly small values of $r$; in particular, the region corresponding to $n_s\simeq 0.967$ yields $r\lesssim 10^{-7}$.
This is a powerful statement in terms of upcoming precision measurements --- if Planck measures a non-negligible value for $r$, inflation models of this type will be convincingly ruled out.

It is interesting to note that the vacuum energy scale $V_0^{1/4}=\sqrt{\kappa}M$ in these models takes on values that can be substantially smaller than the typical scale $\sim 10^{16}$~GeV.  As a direct consequence, the $G$-symmetry breaking scale $M$ turns out to be at most $\sim 5\times 10^{15}$~GeV for $\mathcal{N}=1$ (i.e. for a U(1) model).  In the flipped SU(5) model, however, $M$ can be significantly larger; as can be seen in Figs.~\ref{rnsk_p}(b),~\ref{rnsk_z}(b) and~\ref{rnsk_n}(b), this model leads to $M\lesssim 7$--$8\times 10^{15}$~GeV.  This value acts as an upper bound for $a=0$,~$-1$ and for $|M_S|\gtrsim 10^7$~GeV in the $a=+1$ case.  For $a=1$, $|M_S|\lesssim 10^7$~GeV, $M$ can be substantially larger than $10^{16}$~GeV (see Fig.~\ref{MkMS}(a)).  However, this region of parameter space also corresponds to very low, possibly unnatural values of $\kappa$, as can be seen in Fig.~\ref{MkMS}(b).

It has recently been shown that similar scales, $M\sim 6$--$7\times 10^{15}$~GeV, are exhibited in the breaking of flipped SU(5) with threshold corrections taken into account \cite{Li:2009fq}.  In such models, a reduced proton lifetime of order $\sim 10^{34}$~yr is obtained in the channel $p\rightarrow \pi^0 e^+$, lying within the range to be probed by future planned experiments such as Hyper-K and DUSEL.  According to Fig.~\ref{MkMS}(a), then, a SUSY hybrid model involving a negative soft mass-squared term of order $|M_S|\sim 10^{9}$--$10^{10}$~GeV appears to be in comfortable agreement with both WMAP and proton decay predictions.  On the other hand, we find that $M\lesssim 1$--$2\times 10^{15}$~GeV for $|M_S|=0$, with $a=-1$ and $n_s\approx 0.967$.  Choosing larger $|a|$ values in this case increases $M$, yet this enhancement is not enough even for unnaturally large $|a|$ (e.g. $M\simeq 3.6\times 10^{15}$~GeV for $a=-100$).  Thus we see that $M_S^2 < 0$ is needed for consistency with proton decay considerations in a flipped SU(5) model.

\section*{Summary}

In summary, we have explored the consequences of including the contribution of a sizable soft SUSY-breaking mass-squared term in the hybrid inflationary potential.  As long as the physical inflaton mass is canonical, this soft term may be accompanied by a negative sign.  If the magnitude of this soft mass $|M_S|$ is in an intermediate range, the WMAP5 central value of the spectral index $n_s\simeq 0.967$ (for $r\sim 0$) may be obtained.  The presence and sign of a soft linear term can also play a significant role; as this term becomes smaller and then negative, smaller values of $|M_S|$ lead to good agreement with the WMAP5 data.  If these minimal models are embedded within a flipped SU(5) GUT, a symmetry breaking scale $M\sim 8\times 10^{15}$--$2\times 10^{16}$~GeV, which is consistent with a scalar spectral index $n_s = 0.96$--$0.97$, leads to predictions of the proton lifetime of order $10^{34}$--$10^{36}$~yrs.  The model discussed here can be extended to other supersymmetric GUTs such as SU(5) and SO(10).

\section*{Acknowledgments}
We thank Nefer {\c S}eno$\breve{\textrm{g}}$uz for valuable discussions. This work is
supported in part by the DOE under grant No.~DE-FG02-91ER40626, by the University of Delaware competitive fellowship (M.R.), and by NASA and the Delaware Space Grant Consortium under grant No.~NNG05GO92H (J.W.).

\pagebreak

\pagebreak

\pagebreak

\end{document}